\documentclass[referee]{raa}
\pdfoutput=1
\usepackage{graphicx,times}
\usepackage{natbib}
\usepackage{amssymb,amsmath}
\usepackage{braket}
\bibpunct{(}{)}{;}{a}{}{,}
\usepackage{graphicx}
\usepackage{subcaption}

\usepackage[pagebackref=true]{hyperref}

\begin{document}

   \title{The Parameter Dependence of $\mathbf{n_{s}}$ and $\mathbf{r}$ of the Scalar Power Spectrum during Single-Field Slow-Roll Inflation: A Comparative Study of Inflationary Potentials}

 \volnopage{ {\bf 20XX} Vol.\ {\bf X} No. {\bf XX}, 000--000}
   \setcounter{page}{1}

   \author{Guanqiao Liu} 
   

   \institute{ School of Physics and Astronomy, Sun Yat-sen University, Zhuhai, China; {\it liugq9@mail2.sysu.edu.cn}\\
\vs \no
}

\abstract{Inflation in cosmology is a specific stage preceding the Big Bang, aimed at solving both old background problems and new perturbation issues. Single-field inflation is a candidate to illustrate the picture of the initial universe, and various potential functions lead to different scenarios during the inflationary stage. This paper introduces two essential parameters: the spectral index and the tensor-to-scalar ratio detected from the initial power spectrum, derived from the action of  the scalar field and using approximation that the potential is flat. A brief overview of the origins of Starobinsky Inflation, Chaotic Inflation, Small Field Inflation, and Natural Inflation is also presented, along with their mathematical representations. Finally, the results derived from various inflation models regarding the index and ratio are tested using the Planck data, and the deviations in each model are analyzed.
}

   \authorrunning{Guanqiao Liu}            
   \titlerunning{Comparative Study on Parameter Dependence of $\mathbf{n_{s}}$ and $\mathbf{r}$}  
   \maketitle

%
\section{Introduction}
Today, people investigates correlations of matter and radiation in the sky, specifically in the context of large-scale structure as well as the cosmic microwave background, which reflect the distributions of them respectively. Various methods and detectors are employed to explore these correlations, such as Planck~\cite{aghanim2018planck}, DESI~\cite{collaboration_desi_2016}, and Euclid~\cite{laureijs2011euclid}. These projects focus on specific objectives at large scales and produce extensive datasets. The data generated can enhance the understanding of both the primordial universe and its subsequent evolution, advancing cosmology into a precision science. However, to extract the most information from the data, it is essential to identify the signals to look for and the models to employ.

In the initial cosmological theory, several problems remain unexplained, including the curvature problem, horizon problem, phase coherence problem, and scale invariance problem, among others. The first two issues are classic background problems, well known for over 40 years, while the others represent new perturbation problems identified with advancements in cosmological detectors. These challenges prompted cosmologists to assume an inflationary process before the Big Bang. The classical background problems arise from the initial condition constraints of the standard cosmology model, while the perturbation issues are related to the origin of primordial fluctuations. Inflationary theory naturally resolves these contradictions through an early exponential expansion: it smooths out spatial curvature, expands the causally connected region, and provides a dynamical mechanism for a scale-invariant spectrum of perturbations. This model has successfully explained most of the problems, and the computational framework for primordial perturbations has been established by a series of pioneering works, including Bardeen's gauge-invariant formulation~\cite{bardeen1983spontaneous-2003-5}, Mukhanov's theory of quantum perturbations~\cite{mukhanov1981quantum2003-11}, and Starobinsky's predictions of gravitational waves~\cite{starobinsky1982dynamics2003-12}, which can be tested using the detectors mentioned earlier.

Among the various inflation models, single-field inflation is a candidate to describe the picture of the initial universe, and a prediction of this model is derived here. In \cite{maldacena_non-gaussian_2003,paolo_creminelli_single-field_2004}, it was shown that this nice model leads to a distinct prediction for the perturbation signal. This essay primarily utilizes the slow-roll assumption to simplify the complex models and focuses exclusively on scalar fluctuations.

To illustrate the magnitude of perturbations, various functions are employed, such as the power spectrum and the bispectrum. These functions reveal correlations of matter across different regions of the sky at various scales, making them some of the most essential signals to detect and predict. Here, the focus is on the power spectrum, the simplest among these functions, and the changes in the spectral index under different models are calculated.

In single-field inflation, different potential functions can be chosen, such as chaotic inflation \cite{LINDE1983177}, Starobinsky inflation \cite{starobinsky1980new}, and natural inflation \cite{freese1990naturalmodel}. These models are based on different assumptions and directly influence the slow-roll parameters, which have a distinct relationship with the amplitude of spectrum we detected before. Thus, an appropriate model is selected to demonstrate the power spectrum today. Additionally, understanding the details of evolution during inflation is essential for comprehending the time interval of inflation, as single-field inflation predicts possible outcomes regarding the expansion process in the early universe.

In this paper, cosmological fluctuations are discussed under different inflationary potential functions, and the initial power spectrum is analyzed. The focus is on the influence of $n_s$ and $r$, two essential parameters of power spectrum, aiming to select an appropriate model that can explain the data from today’s detectors. Additionally, the time dependence of field of the background and the power spectrum in two-dimensional space is demonstrated under various potential functions, with results visualized using numerical methods. Furthermore, simple constraints on the parameters occurred in these potential functions are provided based on observations from Planck and other detectors, with the goal of assessing the accuracy of these models. Units are used in which \(\hbar = c =  1\), and the Planck mass is defined as \(M_{\text{Pl}} = (8\pi G_N)^{-1/2}\).

\section{Scalar Power Spectrum}

In cosmology, different types of perturbations are explored, including scalar, vector, and tensor perturbations. For instance, when considering the temperature fluctuations detected by the CMB, scalar perturbations are utilized to illustrate these fluctuations. Meanwhile, tensor perturbations are typically employed to represent primordial gravitational waves. This paper focuses solely on the simplest density perturbation and briefly discusses the ratio of the amplitudes of tensor and scalar perturbations to test the predictions of the models.

\subsection{Correlation Function and Scalar Power Spectrum}

The density perturbation is quite complex, and a comprehensive understanding of all the local details is not feasible. Instead, statistical methods are employed to comprehend the distribution of matter and radiation in our universe.

Assuming that the universe has an approximate matter density \(\overline{\rho}(t)\), and that the density perturbation \(\delta\rho(\textbf{x},t)\) is much smaller than \(\overline{\rho}(t)\), a dimensionless quantity is defined:

\begin{equation}
    \delta(\textbf{x},t) = \frac{\delta\rho(\textbf{x},t)}{\overline{\rho}(t)}
    \label{2.1}
\end{equation}

From the cosmological principle, several properties of \(\delta(\textbf{x},t)\) can be derived. Firstly, the spatial average of \(\delta\) vanishes:

\begin{equation}
\langle \delta(\mathbf{x}, t) \rangle = 0
\label{2.2}
\end{equation}

When calculating the correlation between two points in the sky, the correlation function is obtained:

\begin{equation}
\xi(r, t) = \left\langle \delta(\mathbf{x}, t) \delta(\mathbf{y}, t) \right\rangle
\label{2.3}
\end{equation}

Here, $\xi$ is a function of $r$, the distance between two points in the sky. Because the assumption that our universe is statistically same everywhere and at any direction, the correlation function has no relation with any point and the direction of the vector $r$.   This illustrates that this statistic cannot differ at special points or directions and only depends on the relative position.

Next, this essential function is translated into momentum space, where the evolution can be explored more directly. It can be defined as:

\begin{align}
\langle \delta(\mathbf{k}, t) \delta(\mathbf{k'}, t) \rangle
&= \int d^3x \, d^3y \, e^{i\textbf{k}\cdot\textbf{x}+i\textbf{k}\cdot\textbf{y}} \langle \delta(\mathbf{x}, t) \delta(\mathbf{y}, t) \rangle \notag\\
&=  \int d^3r \, (2\pi)^3 \delta^3_D(\textbf{k+k'})e^{i\textbf{k}\cdot\textbf{r}} \xi(r,t)
\label{2.4}
\end{align}

Here the delta function is written as \(\delta_D\), and the integral representation without the delta function and the constant is the definition of power spectrum:

\begin{equation}
P(k, t) = \int d^3r \, \xi(r, t)e^{i\textbf{k}\cdot\textbf{r}} 
\label{2.5}
\end{equation}

Since the correlation function \(\xi(r)\) can be measured using various detectors, the expression for the spectrum can be illustrated through straightforward observations. Thus, the inflationary models can be tested with this function. However, the form of the power spectrum also influences the discussion. Therefore, the power-law form is adopted:
\begin{equation}
P(k) = Ak^{n_s}
\label{2.6}
\end{equation}
The exponent \(n_s\) is the spectral index that we can detect. The expected spectral index in classical cosmology and the deviations from it under different models will be examined. \(A\) is the amplitude, which provides constraints on the parameters in the potential function.

\subsection{Initial Perturbation of Scalar Field}

In the inflationary stage, the Hubble parameter \(H\) is almost a constant, and the scale factor $a$ increases exponentially as \(a \sim e^{H_{\text{inf}} t}\). To explore the development and transformation of quantum perturbations in the context of such spacetime, a scalar field \(\phi\) is introduced, which drives inflation and experiences initial fluctuations. The action of \(\phi\) is given by:

\begin{equation}
    S = \int d^3x \, dt \, a^3(t) \left[ \frac{1}{2} \dot{\phi}^2 - \frac{1}{2a^2(t)} \nabla \phi \cdot \nabla \phi - V(\phi) \right]
\label{2.7}
\end{equation}

Considering the simplest situation where \(V(\phi) = \text{const}\), which is equivalent to a cosmological constant, the dynamics of \(\phi\) can be derived by taking the Euler-Lagrange equation in field theory:

\begin{equation}
    \frac{\partial L}{\partial \phi} - \partial_\sigma \left( \frac{\partial L}{\partial (\partial_\sigma \phi)} \right) = 0
    \label{2.8}
\end{equation}

In general, when we discuss the evolution of the background field, using the form in the momentum space is usually convenient. The equation of scalar field in momentum space is derived as
\begin{equation}
    \frac{d^2 \phi_k}{dt^2} + 3 H_{\text{inf}} \frac{d\phi_k}{dt} + \frac{k^2}{a^2} \phi_k = 0
    \label{2.9}
\end{equation}

Notice that the equation above resembles that of a classical harmonic oscillator with friction, including a factor \(a\)developing with the time. To remove the friction and the scale factor, the analysis proceeds in conformal time \(d\tau = \frac{dt}{a(t)}\) and a new field is defined as \(\tilde{\phi_k} = -\phi_k / (H_{\text{inf}} \tau)\). A concise equation can then be derived:

\begin{equation}
    \frac{d^2 \tilde{\phi_k}}{d\tau^2} + \left(k^2 - \frac{2}{\tau^2}\right) \tilde{\phi_k} = 0
    \label{2.10}
\end{equation}

Here, \(\tau = -1/(a H_{\text{inf}})\) and \(\tau \in (-\infty, 0)\), illustrating the behavior as \(\tau \to 0\). This equation is solved by considering boundary conditions. Firstly, as \(\tau \to -\infty\), the equation behaves like a classical harmonic oscillator, and the solution is:

\begin{equation}
    \tilde{\phi_k} = A e^{\pm ik\tau} 
    \label{2.11}
\end{equation}

When \(\tau \to 0^-\), Eq. (\ref{2.10}) becomes:

\begin{equation}
    \frac{d^2 \tilde{\phi_k}}{d\tau^2} - \frac{2}{\tau^2} \tilde{\phi_k} = 0 
    \label{2.12}
\end{equation}

Assuming that the solution has a power-law form \(\tilde{\phi_k} = A k^n\), where \(n\) is the index, it can be calculated that \(n = 2\) or \(n = -1\). From the two boundary conditions, the final solution can be expressed as:

\begin{equation}
    \tilde{\phi_k} = (C_1 + C_2 \tau^2 + C_3 \tau^{-1}) e^{\pm ik\tau} 
    \label{2.13}
\end{equation}

Substituting Eq. (\ref{2.13}) into Eq. (\ref{2.10}) allows for the derivation of:

\begin{equation}
    \tilde{\phi}_{k} = \alpha e^{-ik\tau} \left(1 - \frac{i}{k\tau}\right) + \beta e^{ik\tau} \left(1 + \frac{i}{k\tau}\right)
    \label{2.14}
\end{equation}

The solutions above contain well-known functions: the spherical Bessel functions. Eq. (\ref{2.14}) can be rewritten using spherical Bessel functions \(j_{1}(x)\) and \(n_{1}(x)\):

\begin{equation}
    \tilde{\phi}_k = k\tau \left( A j_{1}(k\tau) + B n_{1}(k\tau) \right).
    \label{Bessel Function}
\end{equation}

In Figure 1, it is evident that the initial fluctuation in the universe resembles the behavior of a classical harmonic oscillator until \(\tau \to -10\). After that point, the amplitude of \(\phi\) increases and gradually accelerates, which can be attributed to the expansion of the universe. However, when \(k\tau \sim 1\), the signal cannot be transmitted from different parts of the oscillator because the characteristic length approaches the cosmological horizon. Thus, this initial quantum perturbation exits spacetime and becomes part of the background, making further analysis of its evolution unnecessary.
\begin{figure}[h]
    \centering
    \includegraphics[width=1\linewidth]{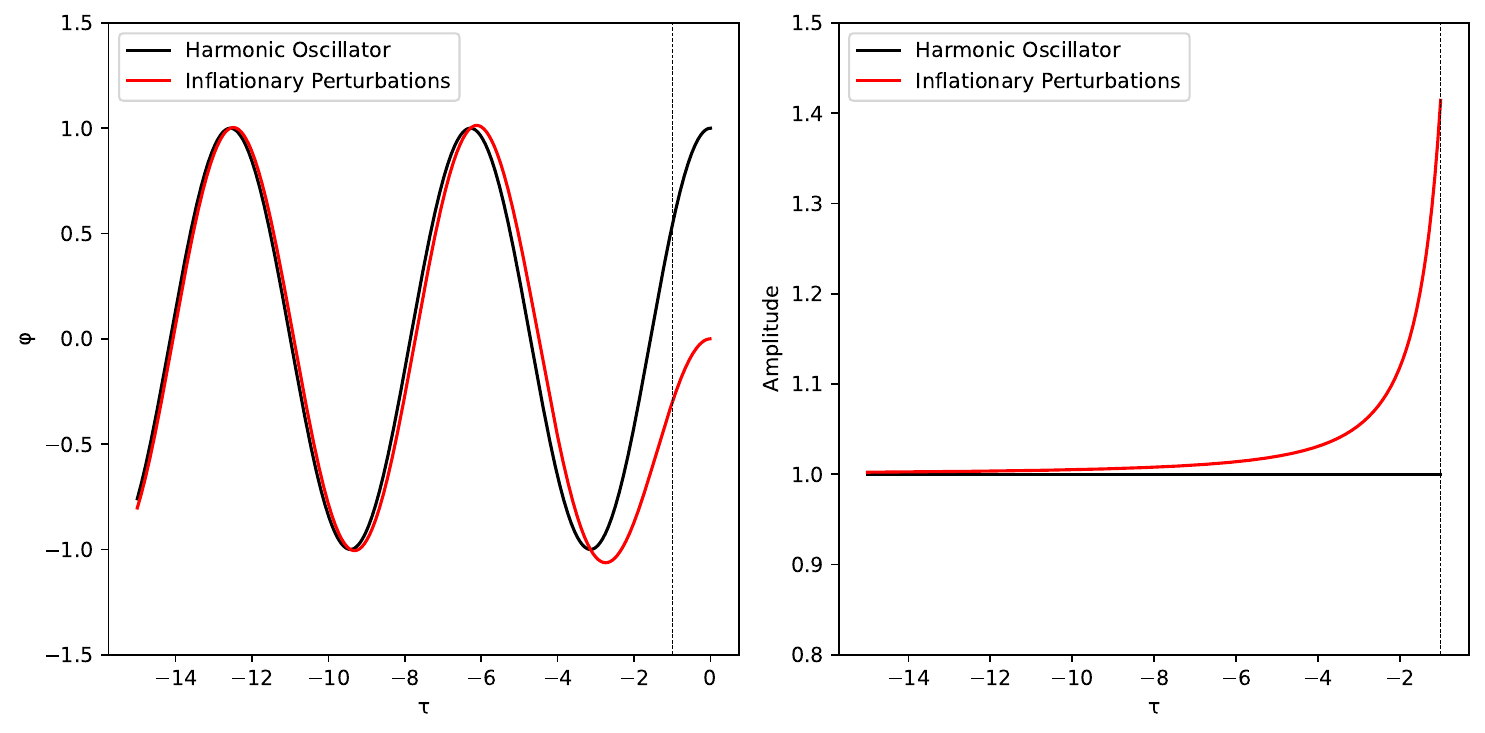}  
    \caption{The left picture illustrates the development of \(\phi\), while the right one demonstrates the amplitudes changing with the conformal time \(\tau\) compared to the classical harmonic oscillator. Here \(A = k = 1\).}
\end{figure}

\subsection{Harrison-Zel'dovich Spectrum}

In Section 2.2, the fluctuation was solved using classical methods, but a specific representation for the amplitude of \(\phi\), which illustrates the power spectrum, was not confirmed. With aim to predict the power spectrum, the time-dependent harmonic oscillator in quantum mechanics is considered.

In quantum mechanics, annihilation and creation operators are typically introduced, which can be written as:

\begin{equation}
    \hat{a} = \sqrt{\frac{\omega}{2}} \, \hat{q} + i \sqrt{\frac{1}{2 \omega}} \, \hat{p} \quad \text{and} \quad \hat{a}^\dagger = \sqrt{\frac{\omega}{2}} \, \hat{q} - i \sqrt{\frac{1}{2 \omega}} \, \hat{p}
    \label{2.15}
\end{equation}

Here, position and momentum can be expressed inversely. Different energy states are

\begin{equation}
    \hat{a}\ket{m} = \sqrt{m}\ket{m-1} \quad \text{and} \quad \hat{a}^\dagger\ket{m} = \sqrt{m+1}\ket{m+1}
    \label{def of a}
\end{equation}

The average and variance of \(\hat{q}\) can be calculated in this basis, focusing only on the ground state:

\begin{align}
   &\langle 0|\hat{q}|0 \rangle = \sqrt{\frac{1}{2\omega}} \langle 0|(\hat{a} + \hat{a}^\dagger)|0 \rangle = 0 \label{2.17} \\ 
   &\langle 0 | \hat{q}^2 | 0 \rangle = \frac{1}{2\omega} \langle 0 | (\hat{a} + \hat{a}^\dagger)^2 | 0 \rangle = \frac{1}{2\omega} \label{2.18}
\end{align}

Thus, \(\langle\phi^{2}\rangle = \frac{\hbar}{2m}\) is obtained. The amplitude of a harmonic oscillator in quantum mechanics can now be derived. However, solving the oscillator equation with a time-dependent frequency requires discussing the evolution of position in a useful picture that named after Heisenberg.

In the Heisenberg picture, the operators change over time. Taking \(\hat{a}(\tau_{0})\) and \(\hat{a}^\dagger(\tau_0)\) as time-independent bases, the operator \(\tilde{\phi}_{k}(\tau)\) can be expressed as:

\begin{equation}
    \hat{\phi}_k(\tau) = v_k(\tau) \hat{a}_k(\tau_0) + v_k^*(\tau) \hat{a}_k^\dagger (\tau_0)
    \label{2.19}
\end{equation}

where \(v(\tau)\) is the coefficient, which varies with time. This coefficient needs to fulfill the equation describing inflation perturbations. (\ref{2.10}). Thus, the solution for \(v(\tau)\) is the same as in (\ref{2.14}).

In quantum mechanics, the uncertainty principle can be expressed as:

\begin{equation}
    [ \hat{q}, \hat{p}] = i \label{2.20}
\end{equation}

where \(\hat{p} = \frac{d\hat{q}}{dt}\). This relationship is applied to \(v(\tau)\) with the assumption that \(\beta\) in (\ref{2.14}) is 0, leading to:

\begin{equation}
    v_k \dot{v_k^{*}} - \dot{v_k} v_k^{*} = 2\alpha^2 ik = i \label{2.21}
\end{equation}

where \(\alpha = \sqrt{\frac{1}{2k}}\).

The final result for \(v(\tau)\) is:

\begin{equation}
    v_k(\tau) = \sqrt{\frac{1}{2k}} e^{-ik\tau} \left(1 - \frac{i}{k\tau}\right) \label{2.22}
\end{equation}

Although the fluctuation of \(\tilde{\phi}_k\) is given by the magnitude of \(v_k\), the original field \(\phi_k = -H_{\text{inf}} \tau \tilde{\phi_k}\) is of greater interest. The perturbations $\phi_k$ are 

\begin{equation}
    \left\langle \hat{\phi}_{k} \hat{\phi}_{k}^\dagger \right\rangle = \frac{H_{\mathrm{inf}}^2}{2k} \left( \frac{1}{k^2} + \tau^2 \right) \label{2.23}
\end{equation}

At early times, the perturbations of the field are large, but they exit the horizon as \(k\tau \to -1\). At subsequent times, the way perturbations vary with k is expressed as:

\begin{equation}
    P_{\phi} = \lim_{k\tau \to 0^{-}} \langle \hat{\phi}_{k} \hat{\phi}_{k}^\dagger \rangle = \frac{H_{\text{inf}}^2}{2k^3} \label{2.24}
\end{equation}

It is frequently beneficial to utilize the dimensionless power spectrum as:

\begin{equation}
    \Delta_{\phi} = \frac{4\pi k^3 P_{\phi}}{(2\pi)^3} = \frac{H_{\text{inf}}^2}{(2\pi)^2} \label{2.25}
\end{equation}

This is the famous Harrison-Zel'dovich spectrum, which closely resembles the real power spectrum. It implies the scale invariance of the correlation function because:

\begin{align}
    \langle \phi(\lambda \mathbf{x}) \phi(\lambda \mathbf{y}) \rangle 
    &= \int d^3k \int d^3k' \frac{1}{(2\pi)^6} e^{-i(\mathbf{k} \cdot \lambda \mathbf{x} - \mathbf{k'} \cdot \lambda \mathbf{y})} \langle \phi(\mathbf{k}) \phi(\mathbf{k}') \rangle \nonumber\\
    &= \int \frac{d^3k}{(2\pi)^3} e^{-i \mathbf{k} \cdot \lambda (\mathbf{x} - \mathbf{y})} P_{\phi}(k) \nonumber\\
    &= \int \frac{d^3(\lambda k)}{(2\pi)^3} e^{-i \mathbf{k} \cdot \lambda (\mathbf{x} - \mathbf{y})} \frac{H_{\text{inf}}^2}{2(\lambda k)^3} \nonumber\\
    &= \langle \phi(\mathbf{x}) \phi(\mathbf{y}) \rangle \label{2.26}
\end{align}

Here, the result does not change with the scale of the two points because \(P_{\phi} \propto k^{-3}\). However, the power spectra observed are not always for massless particles; real particles always obey the Poisson equation:

\begin{equation}
    \nabla^2 \delta \phi = \frac{(1 + 3w)}{2M_{\text{Pl}}^2} \overline{\rho} a^2 \delta \label{2.27}
\end{equation}

where \(w = \frac{p}{\rho}\).

In momentum space, it is straightforward to see that \(k^2 \delta \phi_k \propto \delta_k\), indicating that the matter power spectrum is always proportional to \(k\) under the assumption of scale invariance. Consequently, the spectral index is equal to one.
\begin{figure}[htbp]
    \centering
    \subfloat[n=2]{\includegraphics[width=0.45\textwidth]{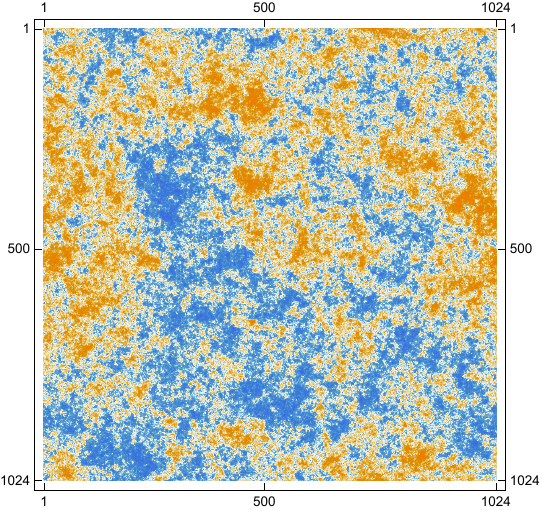}}\label{fig:image1}
    \hfill
    \subfloat[n=3]{\includegraphics[width=0.45\textwidth]{"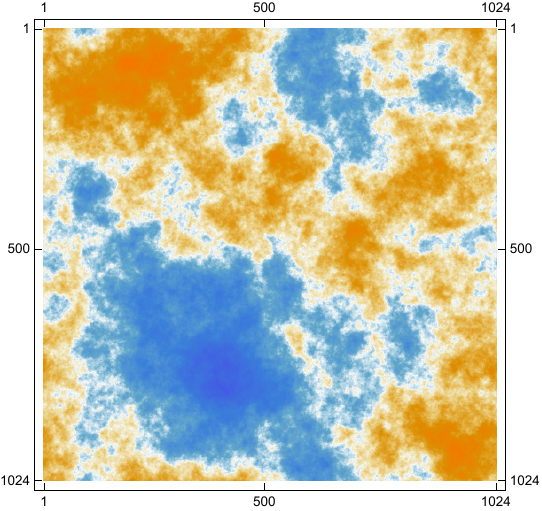"}\label{fig:image2}}
    \caption{These two figures illustrate the distributions of power spectra in two-dimensional space for \(n=2\) and \(n=3\), where \(P = \frac{C}{k^n}\). \(n=2\) indicates scale invariance.}
    \label{fig:two_images}
\end{figure}

In Figure 2, it is clear that when the power spectrum is scale-invariant, there are not many special structures on large scales. However, the blue region appears in the southeast of the figure, while the southwestern part is red when \(n > 2\). Current observations indicate that \(n_s\) is slightly smaller than 1, suggesting that the true condition lies between (a) and (b), closer to scale invariance.\footnote{The distribution images here are generated by Mathematica script written by Garrett Goon. The details can be found at \url{https://garrettgoon.com/gaussian-fields/}.}

This section discusses the classical results of the power spectrum, noting that if scale invariance were considered a fundamental principle, \(n_s\) would always equal one. However, observations from Planck indicate a small deviation from unity for $n_s$. To clarify this deviation, the single-field inflation model will be introduced.

An initial power spectrum will be introduced. Instead of the aesthetically pleasing but meaningless one defined by $\phi$ , the real power spectrum, defined by the correlation of matter in the sky, illustrates the fluctuations of the background and exists around the horizon. Using $\delta\phi$ to represent the perturbation of time, specifically $\delta t=\frac{\delta\phi}{\phi}$ , the evolution of the field can be expressed as fluctuations in time. A dimensionless quantity $ \zeta{:=}-H_{\mathrm{inf}}\delta t$ is then introduced. The expression for $ P_\zeta $ is given by

\begin{equation}
    P_{\zeta} = \frac{H^2}{\dot{\phi}^2} \Delta_{\phi} = \frac{H^4}{(2\pi \dot{\phi})^2}
    \label{2.28}
\end{equation}

The index of $k$ in the spectrum above is $n_s$, and large amount of observations and theories are trying to explore this parameter. In Section 4 it is one of the most essential factors to help us choose the proper models and the specific representations of index will be calculated later.
\section{Single-Field Slow-Roll Inflation}

In Harrison-Zel’ dovich Spectrum, it was assumed that \(V(\phi) = \text{const}\), indicating that the potential function does not influence the evolution of the scalar field. A more complex \(V(\phi)\) must be discussed to illustrate the small deviation in the spectral index.

The approximation that the potential is flat can be employed to simplify calculations, which is called slow-roll assumption. Two of the parameters used to describe this assumption can be defined as:

\begin{equation}
    \epsilon:=-\frac{\dot{H}}{H^2} \quad \eta:=\frac{\dot{\epsilon}}{H\epsilon}\label{3.1}
\end{equation}

During the inflationary stage, as long as the matter sector satisfies the Null Energy Condition, \(\epsilon > 0\) (or \(w < -1\)). It is recognized that acceleration requires \(\frac{\ddot{a}}{a} = H^2(1 - \epsilon) > 0\). At this stage, the universe is primarily governed by the cosmological constant, and \(w \approx -1\) indicates that \(0 < \epsilon \ll 1\). Thus, \(\epsilon\) changes significantly during inflation, transitioning from \(\epsilon_0 \ll 1\) to \(\epsilon_{\text{end}} = 1\). Additionally, the second slow-roll parameter can be evaluated around some reference time \(N_*\) using the Taylor expansion of \(\epsilon\):

\begin{equation}
\begin{split}
    \epsilon_{end}-\epsilon_0 & = \left.\ \ \frac{\partial\epsilon}{\partial N}\right|_{N_*}\Delta N+O\left(\left(\Delta N\right)^2\right) \nonumber\\ 
    & =\epsilon\left[\eta\Delta N+O\left(\eta^2\right)\right]
\end{split}
\end{equation}

From the above equation, the requirement that \(\epsilon\) does not change significantly during inflation is given by \(\eta \Delta N_{\text{inf}} < 1\), indicating that the slow-roll approximation needs \(\epsilon, \eta \ll 1\).

Sometimes, another definition is used, referred to as the potential slow-roll parameters:

\begin{equation}
    \epsilon_V := \frac{M_{\text{Pl}}^2}{2} \left( \frac{V'}{V} \right)^2 \quad \eta_V := M_{\text{Pl}}^2 \frac{V''}{V} \label{3.2}
\end{equation}

Using the Friedmann Equation during the inflationary stage, it can be understood that the potential parameters have the same magnitude as the ordinary ones, so \(\epsilon_V, \eta_V \ll 1\) are also valid.

\subsection{Fundamental Equations}

To consider the massless scalar field in a general spacetime, Eq. (\ref{2.7}) needs to be rewritten in GR. The minimally coupled, canonical scalar field action is what we care:

\begin{equation}
    S=-\int d^4x\frac{\sqrt{-g}}{2}\left\{\left[\partial_\sigma\phi\partial^\sigma\phi+2V\left(\phi\right)\right]+M_{Pl}^2R\right\}.
    \label{3.3}
\end{equation}

In General Relativity, the energy-momentum tensor $T_{\mu\nu}$ can be derived from the action, namely

\begin{equation}
    T_{\mu\nu} = \frac{2}{\sqrt{-g}} \frac{\partial S}{\partial g^{\mu\nu}} \label{3.4}
\end{equation}

This equation can be utilized to derive the famous Einstein Equation. The \(T_{\mu\nu}\) in Eq.(\ref{3.3}) can then be calculated as follows:

\begin{equation}
  T_{\mu\nu}=\left[\partial_\mu\phi\partial_\nu\phi-g_{\mu\nu}\frac{1}{2}\partial_\lambda\phi\partial^\lambda\phi\right]-g_{\mu\nu}V\left(\phi\right),
\end{equation}

where the relation \(\delta\sqrt{-g} g_{\mu\nu} \delta g^{\mu\nu}\) is taken into account.

In cosmology, since the scale is vast and the interactions among different objects are nearly negligible, it is often assumed that matter behaves as a relativistic perfect fluid, which is homogeneous and isotropic. If a reference frame is chosen that moves with a point in the fluid, it is straightforward to understand that

\begin{equation}
    T_{\mu\nu} = \text{diag}(\rho, p, p, p)
    \label{3.6}
\end{equation}

Here, \(T_{\mu 0} = 0\) and \(T_{ij} = 0 \, (i \neq j)\) indicate that there are no special directions in the perfect fluid.

From Eq.(\ref{3.6}) and the four-velocity $u^\mu$, $T_{\mu\nu}$ written as the covariant form can be derived as

\begin{equation}
    T^{\mu\nu} = (\rho + p) u^{\mu} u^{\nu} + g^{\mu\nu} p.
    \label{3.7}
\end{equation}

By comparing the two energy-momentum tensors, the following relations can be derived:

\begin{align}
    \rho &= -\frac{1}{2} \partial_{\mu}\phi \partial^{\mu}\phi + V(\phi) \label{3.8} \\
    p    &= -\frac{1}{2} \partial_{\mu}\phi \partial^{\mu}\phi - V(\phi) \label{3.9} \\
    u_{\mu} &= \frac{\partial_{\mu}\phi}{\sqrt{-\partial_{\mu}\phi \partial^{\mu}\phi}} \label{3.10}
\end{align}

If only the perturbations of the homogeneous background are considered, it is useful to assume that the field depends solely on time, so that

\begin{equation}
    \partial_{\mu}\phi \partial^{\mu}\phi = -\dot{\phi}^2 \label{3.11}
\end{equation}

The Friedmann equation and fluid equation describing the evolution of the universe can be rewritten as
\begin{align}
     \frac{1}{2} \dot{\phi}^2 + V(\phi) &= 3H^2 M_{\text{Pl}}^2 \label{3.12} \\
    \ddot{\phi} + 3H \dot{\phi} + V'(\phi) &= 0 \label{3.13}
\end{align}

A useful relation can be obtained by using two equations above and the derivative of Eq.(\ref{3.12}):

\begin{equation}
    -\dot{H} M_{\text{Pl}}^2 = \frac{1}{2} \dot{\phi}^2 \label{3.14}
\end{equation}

Using Eq. (\ref{3.12}), Eq. (\ref{3.14}), a relation between the potential function and inflationary parameters can be established:

\begin{equation}
    V = (3 - \epsilon) H^2 M_{\text{Pl}}^2 \label{3.15}
\end{equation}

Under the assumption that $V',V''\ll\ V$, the correlations between the initial parameters and the potential ones are given by:

\begin{equation}
    \epsilon \approx \epsilon_V \quad \text{and} \quad \eta \approx 4\epsilon_V - 2\eta_V, \label{3.16}
\end{equation}

which illustrates that both types of parameters obey the approximation. Now, the relation between the potential function \(V(\phi)\) and the initial power spectrum can be illustrated using the equations above. However, it is challenging to solve these equations analytically, and the slow-roll approximation can simplify them further.

Firstly, define \(X = \frac{1}{2} \dot{\phi}^2\), leading to the first slow-roll parameter:

\begin{equation}
    \epsilon = -\frac{\dot{H}}{H^2} = \frac{X}{H^2 M_{\text{Pl}}^2} \ll 1. \label{3.17}
\end{equation}

Thus, Eq. (\ref{3.10}) can be simplified as:

\begin{equation}
    3H^2 M_{\text{Pl}}^2 \approx V(\phi) \label{3.18}
\end{equation}

Secondly, the second slow-roll parameter \(\eta\) becomes:

\begin{equation}
    \eta = \frac{\dot{\epsilon}}{\epsilon H} = 2\epsilon + \frac{\dot{X}}{X H}. \label{3.19}
\end{equation}

Since \(\epsilon, \eta \ll 1\), it can be inferred that:

\begin{equation}
    \dot{X} \ll X H \longrightarrow 2\ddot{\phi} \ll \dot{\phi} H, \label{3.20}
\end{equation}

allowing us to neglect the acceleration term \(\ddot{\phi}\) in Eq. (\ref{3.11}):

\begin{equation}
    3H \dot{\phi} \approx -V'. \label{3.21}
\end{equation}

Returning to Eq. (\ref{2.27}), and using Eq. (\ref{3.12}), it can be understood that \(p_\zeta\) is a constant if \(H_{\text{inf}}\) remains stable, which is same as the Harrison-Zel’dovich Spectrum. Thus, the spectral index of the initial power spectrum is \(n_{\zeta} = n_s - 1\), where \(n = \frac{d \ln P_{\zeta}}{d \ln k}\). However, \(P_{\zeta}\) is not a function of \(k\), and a relation between \(k\) and \(\phi\) must be established.

As mentioned in Section 2.3, the quantum fluctuation exits the horizon when \(k\tau = -\frac{k}{a H_{\text{inf}}} \approx 1\) and becomes part of the background. The initial power spectrum is merely a relic of this, obeying the same relation \(k \propto a\), thus:

\begin{equation}
    n_s - 1 = \frac{d \ln P_{\zeta}}{d \ln k} = \frac{a d \ln P_{\zeta}}{da} = \frac{\dot{\phi}}{H} \frac{d \ln P_{\zeta}}{d\phi} \label{3.22}
\end{equation}

Using Eq. (\ref{2.28}), Eq. (\ref{3.17}), and the definitions of parameters above, the final result is:

\begin{align}
    n_s - 1 
    &=-\frac{M_{\text{Pl}}^2 V'}{V} \frac{d \ln (P_{\zeta})}{d\phi} \nonumber\\
    &=-\frac{M_{\text{Pl}}^2 V'}{V} \frac{d}{d\phi} \ln \left( \frac{9 H^2 \times H^4}{4\pi^2 V'^2} \right) \nonumber\\
    &=-\frac{M_{\text{Pl}}^2 V'}{V} \left( 3\frac{V'}{V} - 2\frac{V''}{V'} \right) \nonumber\\
    &=-6\epsilon_V + 2\eta_V \label{3.23}
\end{align}

Finally, during the inflationary stage, two other important parameters are often considered: the e-foldings of expansion and the tensor-to-scalar ratio $r$. The e-foldings is given by:

\begin{equation}
    \Delta N = N_2 - N_1 = \log \left( \frac{a_2}{a_1} \right) \label{3.24}
\end{equation}

It is evident that \(\Delta N\) can be calculated by

\begin{equation}
    N = \int dN = \int H dt = \int \frac{H}{\dot{\phi}} d\phi \approx \int \frac{d\phi}{M_{\text{Pl}} \sqrt{2 \epsilon_V}} = \int d\phi \frac{V}{M_{\text{Pl}}^2 V'},
    \label{eq:integral efolding}
\end{equation}

From current observations, \(\Delta N \in (50, 60)\), which is very useful for constraining the parameters in models.

r is an important number that illustrates distributions of different types of perturbations. Although the general derivation of the tensor power spectrum from our single field assumption is slightly complex, it can be noted that they share the same \(n_s\) in linear theory. The only difference is their amplitudes. In fact, primordial gravitational waves, the source of tensor perturbations, can be described by two modes due to the traceless and transverse conditions. These two modes can be separated by taking the polarization tensors.

\begin{equation}
     \gamma_{ij}(k) = \frac{\sqrt{2}}{M_{\text{Pl}}} \left[ \gamma_{+}(k) e^{+}_{ij}(k) + \gamma_{\times}(k) e^{\times}_{ij}(k) \right], \label{6.25}
\end{equation}

and \(P_{\gamma^{+,\times}}\) has the same form as Eq. (\ref{2.25}) \cite{wang_inflation_2014}. The definition of the tensor spectrum is 

\begin{equation}
     \bar{h}^{ik} \bar{h}^{jl} \langle \gamma_{ij}(\mathbf{k}) \gamma_{kl}(\mathbf{k'}) \rangle = (2\pi)^3 \delta^3(\mathbf{k} + \mathbf{k'}) \frac{2\pi^2}{k^3} P_\gamma \label{3.26}
\end{equation}

Comparing the two equations above, we find

\begin{equation}
    P_{\gamma} = \frac{2H^2}{\pi^2 M_{\text{Pl}}^2}. \label{3.27}
\end{equation}

With Eq. (\ref{2.28}), the tensor-to-scalar ratio is given as

\begin{equation}
    r = \frac{P_{\gamma}}{P_{\zeta}} = 16\epsilon_V. \label{3.28}
\end{equation}

The ratio today are usually detected by analyzing the \textit{\textbf{B}}-mode polarization of the Cosmic Microwave Background (CMB), and the current observational constraint on \(r\) is \cite{PhysRevD.105.083524}

\begin{equation}
    r_{0.002} < 0.032 \quad (95\% \text{ CL}) \label{3.29}
\end{equation}

\subsection{Potential Functions}

Currently, there are many models attempting to describe the phenomenon during the inflationary stage, which are nicely reviewed in \cite{martin_encyclopaedia_2014}. Generally, these models usually have several parameters including the details of initial universe and people often tune parameters to fit the observation result. Models can be classified by the number of parameters. For example, the Starobinsky Inflation model is a zero-parameter model, the Large Field Inflation model and the Natural Inflation model are one-parameter models, while Small Field Inflation has two parameters. This essay will focus on these four models.

Models can also be classified based on the transformation of the field. If \(\Delta\phi < M_{\text{Pl}}\), the models are considered small field theories. For instance, the SFI model mentioned above is a typical small field theory, while the others are large field theories.

\begin{figure}[h]
    \centering
    \includegraphics[width=1\linewidth]{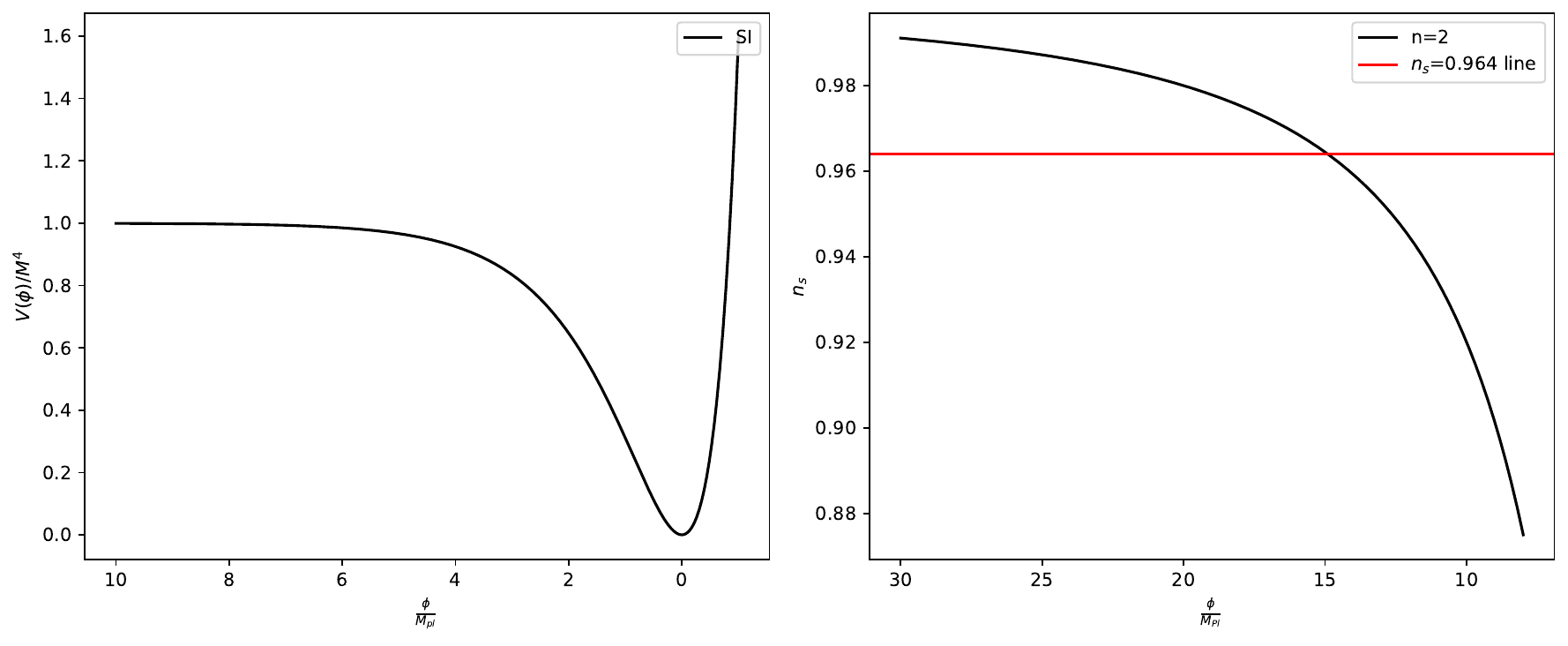}  
    \caption{The left figure shows the plot of \(V(\phi)/M^4\) in Starobinsky inflation. The right figure presents the plot of the spectral index in chaotic inflation. Current observations indicate that the index is approximately 0.964, as depicted by the red line.}
\end{figure}
Firstly, one of the earliest models of inflation was proposed by Alexei Starobinsky in 1980 \cite{starobinsky1980new}. He produced inflation by utilizing a quantum-gravitational effect, examining a FLRW universe populated with quantum fields which are massless, satisfying the covariance in the conformal transformation. This effect modifies the action in General Relativity:

\begin{equation}
    S = \frac{M_{g}^2}{2} \int d^4x \sqrt{-g} f(R), \quad \text{where} \quad f(R) = R + \frac{R^2}{\mu^2}
    \label{3.30}
\end{equation}

Here, \(\mu\) is a parameter that is related to the massless field, which has a mass dimension. From Eq. (\ref{3.30}) and through a complicated derivation \cite{martin_encyclopaedia_2014}, the potential function is explicitly definited as

\begin{equation}
    V(\phi) = M^4 \left( 1 - e^{-\sqrt{\frac{2}{3}} \frac{\phi}{M_{g}}} \right)^2,
    \label{3.31}
\end{equation}

where \(M^4 \equiv \frac{M_g^2 \mu^2}{8}\) and \(M_g \approx M_{\text{Pl}}\) now. In inflationary theories, the parameter $M$ only reflects the whole energy before the Big Bang, resulting in a zero-parameter model.

In Figure 3, it is evident that when \(\phi \ll 2M_{\text{Pl}}\), \(V' M_{\text{Pl}} \ll V\), indicating the slow-roll region. When \(\phi \approx 0\), the universe enters the reheating region, marking the end of inflation.

Secondly, Chaotic Inflation is another early model of inflation, proposed by Linde \cite{LINDE1983177}. It is defined by a simple power-law potential $V\left(\phi\right)\propto\phi^n$, and the index n is the only model parameter.

From Figure 3, the variation of the potential is extremely large, necessitating that \(\phi_0\) be sufficiently large to satisfy the slow-roll approximation. This requirement is clearly illustrated in the right portion of the figure. When \(p = 2\), \(\phi_0\) must exceed \(15M_{\text{Pl}}\) to guarantee that the potential is sufficiently flat. Therefore, it is clear to see that the background field in this model has a large change.

Thirdly, Small Field Inflation is representative of inflation when an extremely flat potential stays at its peak. We can see it in various contexts, such as in \cite{albrecht1982cosmology}, where it is derived with the assumption that it has a background where symmetry is breaking due to the radiation, and this element is also found within frameworks of superstring models \cite{binetruy1986candidates}, supersymmetry \cite{covi2001models}, and others. The potential function for Small Field Inflation is given by

\begin{equation}
    V(\phi) = M^4 \left[ 1 - \left( \frac{\phi}{\mu} \right)^p \right],
    \label{3.32}
\end{equation}

where there are two extra parameters: a parameter describing expectation in vacuum $\mu$ and an index $p$ of the scalar field $\phi$.

In Figure 4, the variations of \(\phi\) and \(V(\phi)\) are both smaller than those of the other models shown here, characterizing it as a small field model. It is observed that this potential generally yields a low tensor-to-scalar ratio.

Finally, People introduce Natural Inflation to tackle the "fine-tuning" issue associated with inflation. This problem arises in many models with numerous parameters, such as \(\mu\) in Small Field Inflation. Furthermore, since the potential \(V(\phi)\) is sufficiently flat during the slow-roll stage, It is frequently contended that this flatness is not stable under quantum corrections unless safeguarded by specific symmetries. In \cite{NI_1992}, Natural Inflation is proposed, where the potential remains smooth because of shift symmetries.

Natural Inflation posits an axion field, resulting in a potential with a periodic structure defined as

\begin{equation}
    V(\phi) = M^4 \left[ 1 + \cos\left( \frac{\phi}{f} \right) \right]
    \label{3.33}
\end{equation}

\begin{figure}[h]
    \centering
    \includegraphics[width=1\linewidth]{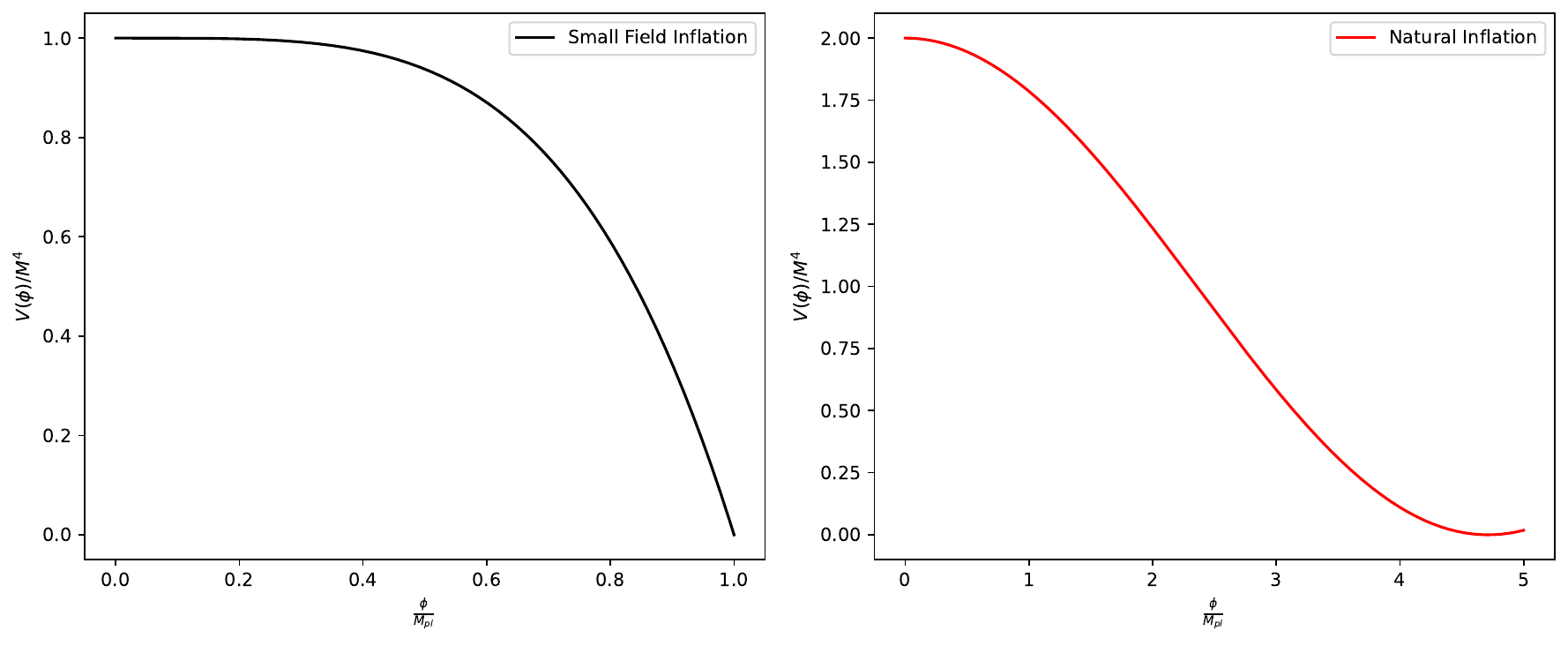}  
    \caption{The figures show the plots of \(V(\phi)/M^4\) in Small Field Inflation and Natural Inflation. In the right figure, \(f/M_{\text{Pl}} = 1.5\), and in the left figure, \(\mu = M_{\text{Pl}}\) and \(p = 4\).}
\end{figure}

From Figure 4, only the change of potential within one period is selected, and we can notice that if the inflation potential keeps flat, \(\phi\) should be close to the extremum of the triangular function, where the derivative is small.

\section{Results \& Discussion}

This section presents the predictions of various models and makes comparisons among them. Visualizing the evolution of $\phi$ and analyzing its transformation trend are often useful, helping us evaluate the time interval during inflation, so including figures of $\phi$ over time is essential for understanding inflation.

Firstly, the expansion fold \(N \in (50, 60)\) and the boundary condition \(\epsilon_V(\phi_{\text{end}}) = 1\) are utilized to derive \(\phi_{\text{ini}}\). As discussed previously, the initial power spectrum is generated when \(\phi = \phi_{\text{ini}}\), so the value of the initial field provides important results. Additionally, the evolution of \(\phi\) is generated by solving the following ordinary differential equation (ODE):

\begin{equation}
    \frac{d\phi}{dt} = -\frac{V' M_{\text{Pl}}}{\sqrt{3V}}.
\end{equation}

This equation can be derived from Eq. (\ref{3.18}) and Eq. (\ref{3.21}). In this paper, the amplitude \(M\) in the potential functions is taken as one to facilitate comparisons of field evolution across different models.

the possible values of two parameters $n_s$ and $r$ are calculated for various models. The calculations reveal deviations between observations and theoretical predictions, allowing for the selection of a more accurate model.

Finally, a comparative study of the models is essential for understanding the drawbacks of each, which aids in refining the theories.

\subsection{Starobinsky Inflation}

\begin{figure}[h]
    \centering
    \includegraphics[width=1\linewidth]{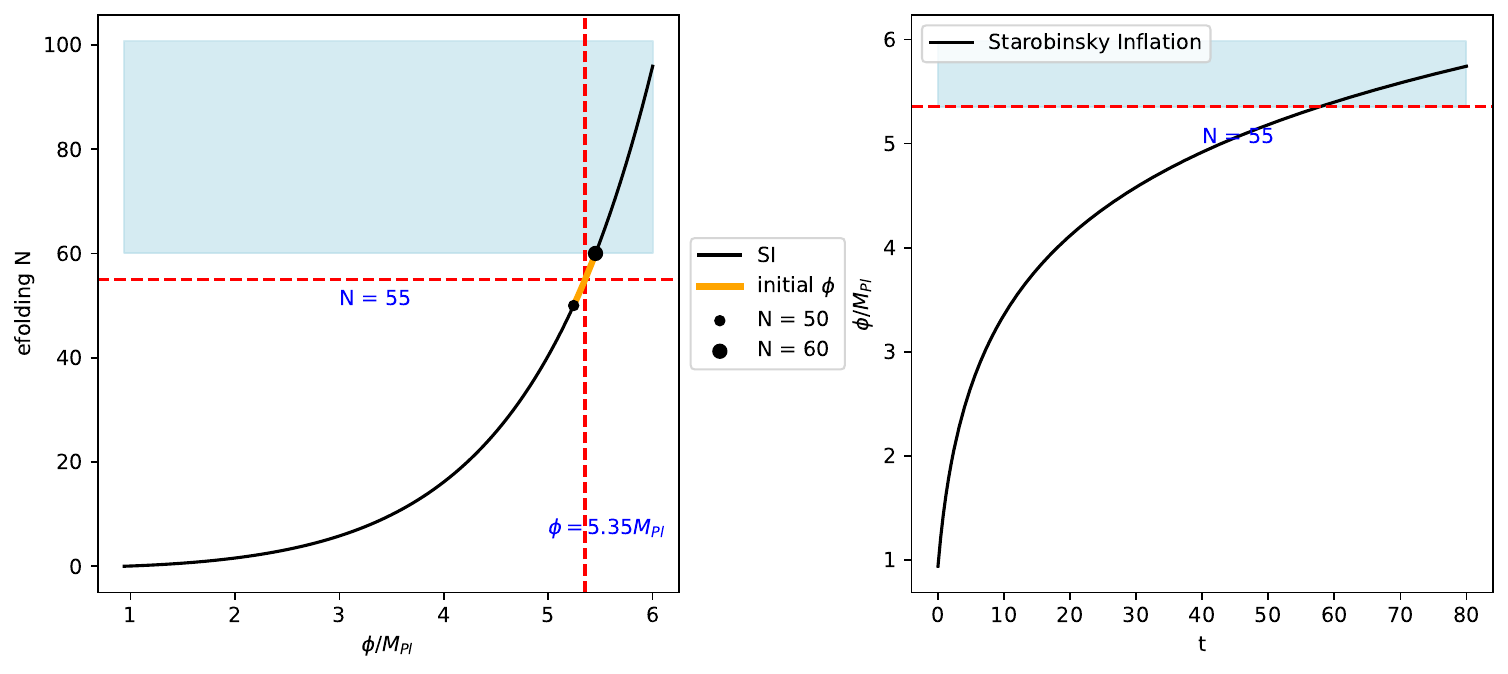}  
    \caption{The left figure illustrates the e-folding varying with the field. The right figure shows the function of \(\phi(t)\), which ceases to change when \(\phi = \phi_{\text{ini}}\).}
\end{figure}

 Figure 5 illustrates the evolution of the e-folding and the scalar field in Starobinsky inflation, where the blue zones indicate forbidden regions. For figures of other models, this style will be maintained.

Using Eq. (\ref{3.31}) and the definition of slow-roll parameters in Eq. (\ref{3.2}), it is straightforward to derive that 

\begin{equation}
     \epsilon_V=\frac{4}{3}[\frac{e^{-2\sqrt{\frac{2}{3}}\frac{\phi}{M_{Pl}}}}{(1-e^{-\sqrt{\frac{2}{3}}\frac{\phi}{M_{Pl}}})^2}]\approx\frac{4}{3}(e^{-2x})\quad,\quad\eta_V=\frac{4}{3}\frac{(e^{-x})-2e^{-2x}}{(1-e^{-x})^2}\approx\frac{4}{3}e^{-x},\label{SIparams}
\end{equation}

where \(x := \sqrt{\frac{2}{3}}\frac{\phi}{M_{Pl}}\) and \(e^{-x}\ll1\) due to the slow-roll approximation.

Although it is typically necessary to calculate \(\phi_{\text{ini}}\) and \(\phi_{\text{end}}\) to derive the final $n_s$ and $r$, in Starobinsky inflation, it is straightforward to obtain these values through the integral result of e-folding \(N\).

Using Eq. (\ref{eq:integral efolding}), the e-folding in Starobinsky inflation is given by

\begin{equation}
    \Delta N = \frac{3}{2}[(e^x-e^{x_{end}})+(x-x_{end})]\approx\frac{3}{2}e^x,
\end{equation}

where \(e^{x_{end}}\approx\sqrt{\frac{4}{3}}\ll e^x\).

Thus, the results for the index and ratio are as follows:

\begin{equation}
    n_s \approx 1-\frac{2}{\Delta N}\approx0.964\quad,\quad r \approx \frac{3}{\Delta N^2}\approx0.001<0.032
\end{equation}

These results fit the observational data very well. Additionally, the right figure in Figure 5 illustrates the evolution of \(\phi\) during inflation. The magnitude of the initial field declines from \(\phi = \phi_{\text{ini}}\), and its rate of variation increases continuously, satisfying the slow-roll approximation \(\epsilon_V \ll 1\) during the beginning and middle of inflation. The horizontal axis of the right figure represents time during inflation, and the intersection point of the red and black plots reflects the duration of this stage.

\subsection{Chaotic Inflation}
\begin{figure}[h]
    \centering
    \includegraphics[width=1\linewidth]{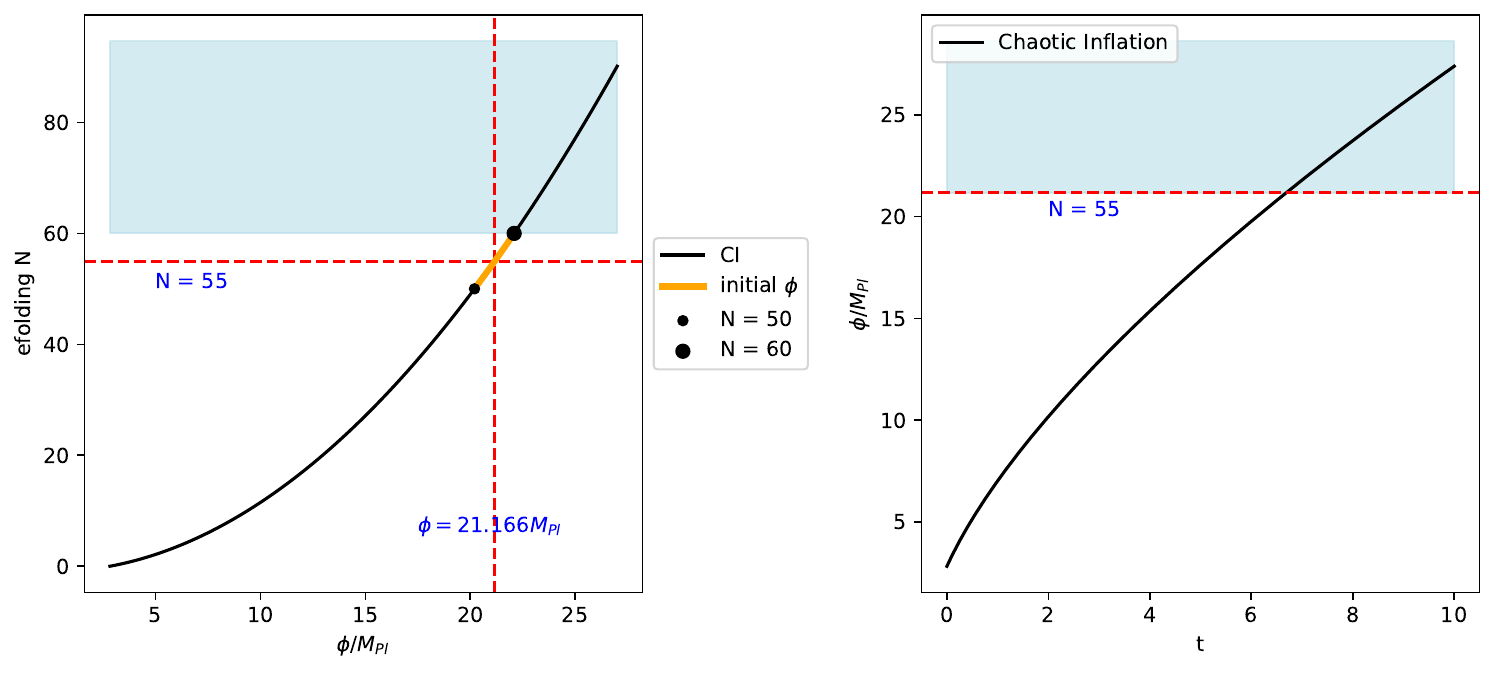}  
    \caption{The left figure illustrates the e-folding varying with the field. The right figure shows the function of \(\phi(t)\), which ceases to change when \(\phi = \phi_{\text{ini}}\).}
\end{figure}

Using power law potential functions, the slow-roll parameters and e-folding are given by

\begin{equation}
    \epsilon_V = \frac{n^2 M_{\text{Pl}}^2}{2\phi^2}, \quad \eta_V = \frac{n(n-1)M_{\text{Pl}}^2}{\phi^2}, \quad N = \frac{\phi^2 - \phi_{\text{end}}^2}{2n} \label{4.5}
\end{equation}

When \(\epsilon_V(\phi) = 1\), it follows that \(\phi_{\text{end}} = \frac{n}{\sqrt{2}}\). Solving Eq. (\ref{4.5}), $n_s$ and $r$ in Chaotic Inflation are determined as
\begin{equation}
    n_s = \frac{4\Delta N - 2n - 3}{4\Delta N + 1} \quad,\quad r = \frac{16n}{4\Delta N + 1}.
\end{equation}

For \(n > 1\), these equations yield \(r > 0.066\), which is significantly greater than the constraint \(r < 0.032\). Conversely, for \(n < 1\), \(n_s > 0.97\), slightly exceeding observational limits, where \(\Delta N\) takes its extremum. Figure 9 illustrates this issue clearly; thus, chaotic inflation is not an appropriate model to describe the early universe. However, power law potentials are often regarded as boundaries. For instance, in Figure 9, the \(r-n_s\) plane is divided into convex and concave areas by the function \(V = M^4 \phi\). In addition, Chaotic Inflation has a large variation of background field and it is easy to infer some general feathers of LFI by analyzing the result here, including a huge deviation on .

\subsection{Small Field Inflation}

\begin{figure}[h]
    \centering
    \includegraphics[width=1\linewidth]{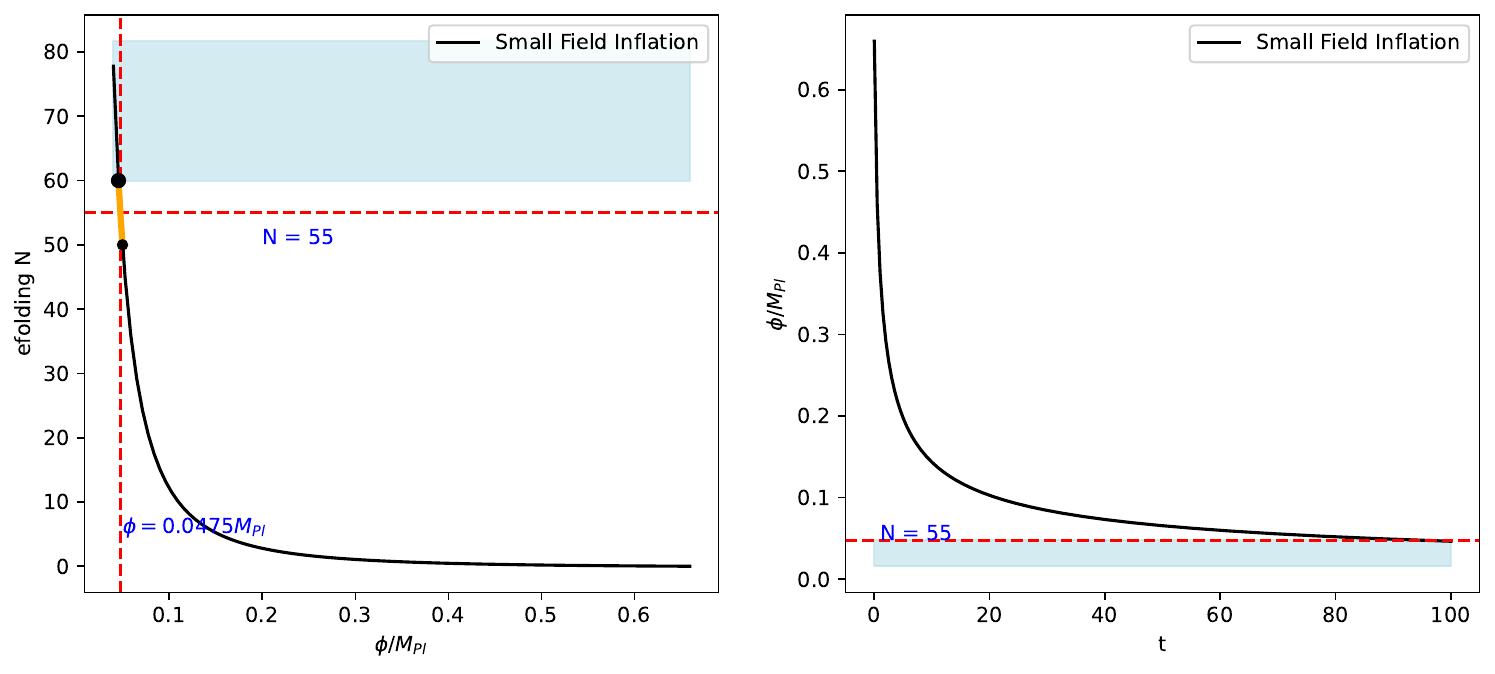}  
    \caption{The left figure illustrates the e-folding varying with the field. The right figure shows the function of \(\phi(t)\), which ceases to change when \(\phi = \phi_{\text{ini}}\), where \(p = 4\) and \(\mu = 1\).}
\end{figure}
In this model, the variation of $\phi$ is less compared to the energy scale $\mu$, and it increases with time during inflation. Using Eq. (\ref{3.32}), the slow-roll parameters are given by

\begin{equation}
    \epsilon_V = \frac{p^2}{2} \left(\frac{M_{\text{Pl}}}{\mu}\right)^2 \frac{x^{2p-2}}{R^2}, \quad \eta_V = p(p-1) \left(\frac{M_{\text{Pl}}}{\mu}\right)^p \frac{x^{p-2}}{R},
\end{equation}

and the index and ratio are

\begin{equation}
    n_s = 1 - p \left(\frac{M_{\text{Pl}}}{\mu}\right)^2 \frac{x^{p-2}}{R^2} \left[x^p (p^2 - 2p) - 2p(p-1)\right], \quad r = 8p^2 \left(\frac{M_{\text{Pl}}}{\mu}\right)^2 \frac{x^{2p-2}}{R^2}.
\end{equation}

Here, \(x := \frac{\phi}{\mu}\) and \(R := \left(1 - \left(\frac{\phi}{\mu}\right)^p\right)\). The representation of \(\Delta N\) should be discussed in different situations. When \(n \neq 2\),

\begin{equation}
 \begin{aligned}
 \Delta N = -\frac{1}{2p} \frac{\mu^2}{M_{\text{Pl}}^2} \left[ -x^2 + x_{\text{end}}^2 + \frac{2}{2-p} \left( x^{2-p} - x_{\text{end}}^{2-p} \right) \right]
 \end{aligned}
\end{equation}

However, when \(p = 2\), the equation above is not applicable, and the integral result is 

\begin{equation}
\Delta N = -\frac{1}{4} \frac{\mu^2}{M_{\text{Pl}}^2} \left[ -x^2 + x_{\text{end}}^2 + 2 \ln \left( \frac{x}{x_{\text{end}}} \right) \right],
\end{equation}

where \(x_{\text{end}}\) satisfies \(\epsilon_V(x_{\text{end}}) = 1\).

Both of these equations cannot be solved analytically in full generality. Thus, numerical solutions for the parameters are evaluated. When \(\mu = 1\) and \(p = 4\), the calculations yield

\begin{equation}
    n_s \approx 0.95, \quad r < 10^{-4},
\end{equation}

as illustrated in Figure 9. The ratio here is very small, but the index of spectrum shows a significant deviation, which cannot be optimized by changing the value of \(\mu\). Therefore, this model is not a suitable candidate when \(p\) is three or four. However, it predicts a small ratio, and with current observational constraints becoming tighter, other small field inflation models may offer better prospects for predicting favorable results for these parameters.
\subsection{Natural Inflation}

\begin{figure}[h]
    \centering
    \includegraphics[width=1\linewidth]{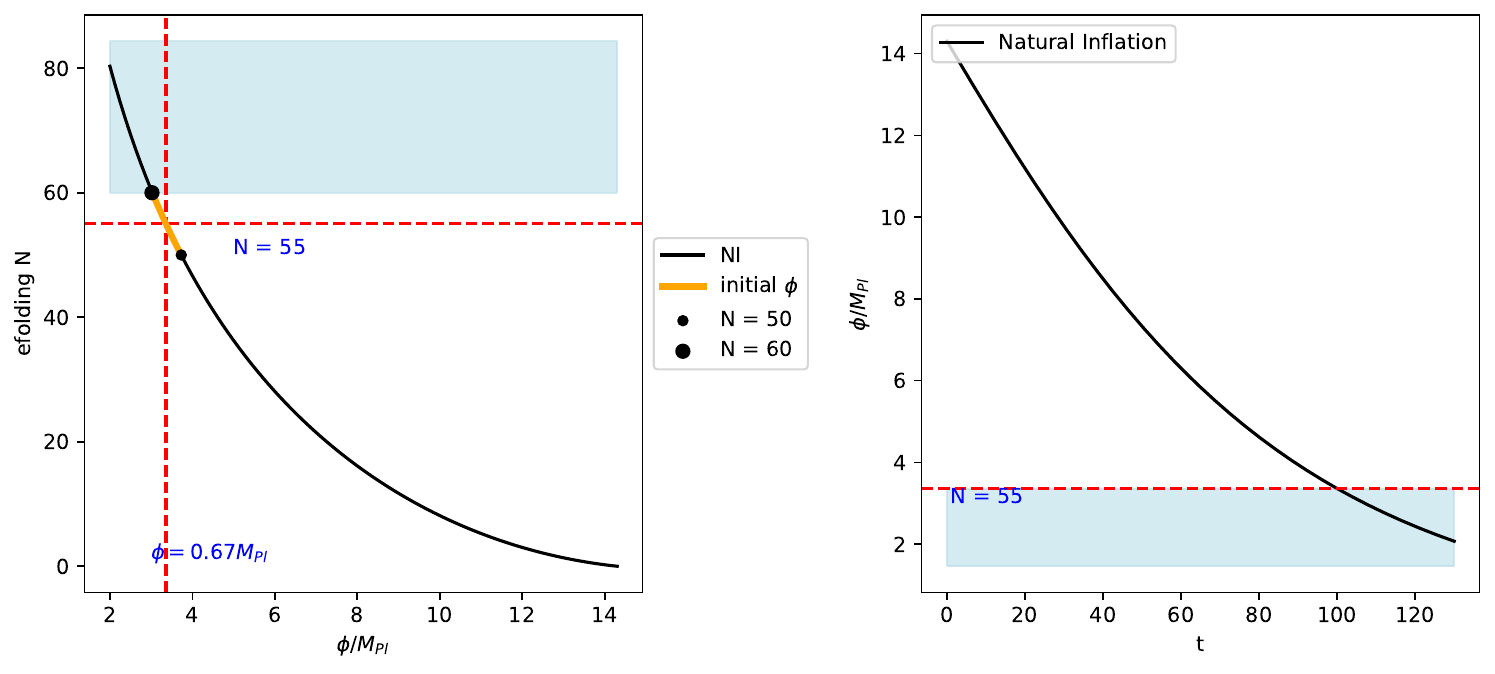}  
    \caption{The left figure illustrates the e-folding varying with the field. The right figure shows the function of \(\phi(t)\), which ceases to change when \(\phi = \phi_{\text{ini}}\), where \(p = 4\) and \(\mu = 1\).}
\end{figure}

In Natural Inflation, the field also increases with time, but the field variation is larger than \(M_{\text{Pl}}\). Using Eq. (\ref{3.33}), the slow-roll parameters are given by 

\begin{equation}
    \epsilon_V = \frac{1}{2f^2}\frac{\sin^2{\frac{\phi}{f}}}{(1+\cos{\frac{\phi}{f}})^2}, \quad \eta_V = -\frac{1}{f^2}\frac{\cos{\frac{\phi}{f}}}{(1+\cos{\frac{\phi}{f}})^2}.
\end{equation}

With the results above, $n_s$ and $ r$ here are

\begin{equation}
    n_s = 1 - \frac{1}{f^2}\frac{3\sin^2{\frac{\phi}{f}} + 2\cos{\frac{\phi}{f}}}{(1+\cos{\frac{\phi}{f}})^2}, \quad r = \frac{8}{f^2}\frac{\sin^2{\frac{\phi}{f}}}{(1+\cos{\frac{\phi}{f}})^2}.
\end{equation}

When \(\epsilon_V(\phi_{\text{end}}) = 1\), inflation ends, leading to

\begin{equation}
    x_{\text{end}} = f \arccos \left( \frac{1 - 2f^2 / M_{\text{Pl}}^2}{1 + 2f^2 / M_{\text{Pl}}^2} \right).
    \label{eq:placeholder}
\end{equation}

Next, the e-folding of expansion is given by

\begin{equation}
    \Delta N = \frac{f^2}{M_{\text{Pl}}^2} \ln \left( \frac{1 - \cos x_{\text{end}}}{1 - \cos x} \right).
    \label{eq:example}
\end{equation}

From Figure 9, when \(f = 5M_{\text{Pl}}\), the parameters are within the \(95\%\) confidence level region, fitting the observational data. However, when \(f = 3M_{\text{Pl}}\), there is a significant deviation in the spectral index. Therefore, in this model, a large energy scale \(f \approx 5M_{\text{Pl}}\) is required, but this introduces the "super-Planck energy scale problem," which may lead to quantum gravitational effects. An effective method to address this issue is to introduce another energy scale \(f_2\) and an additional cosine function \(\cos{\left(\frac{\phi}{f_2}\right)}\), known as Multi-field Inflation \cite{linde1994hybrid}, which may help reduce the energy scale.

\begin{figure}[h]
    \centering
    \includegraphics[width=1\linewidth]{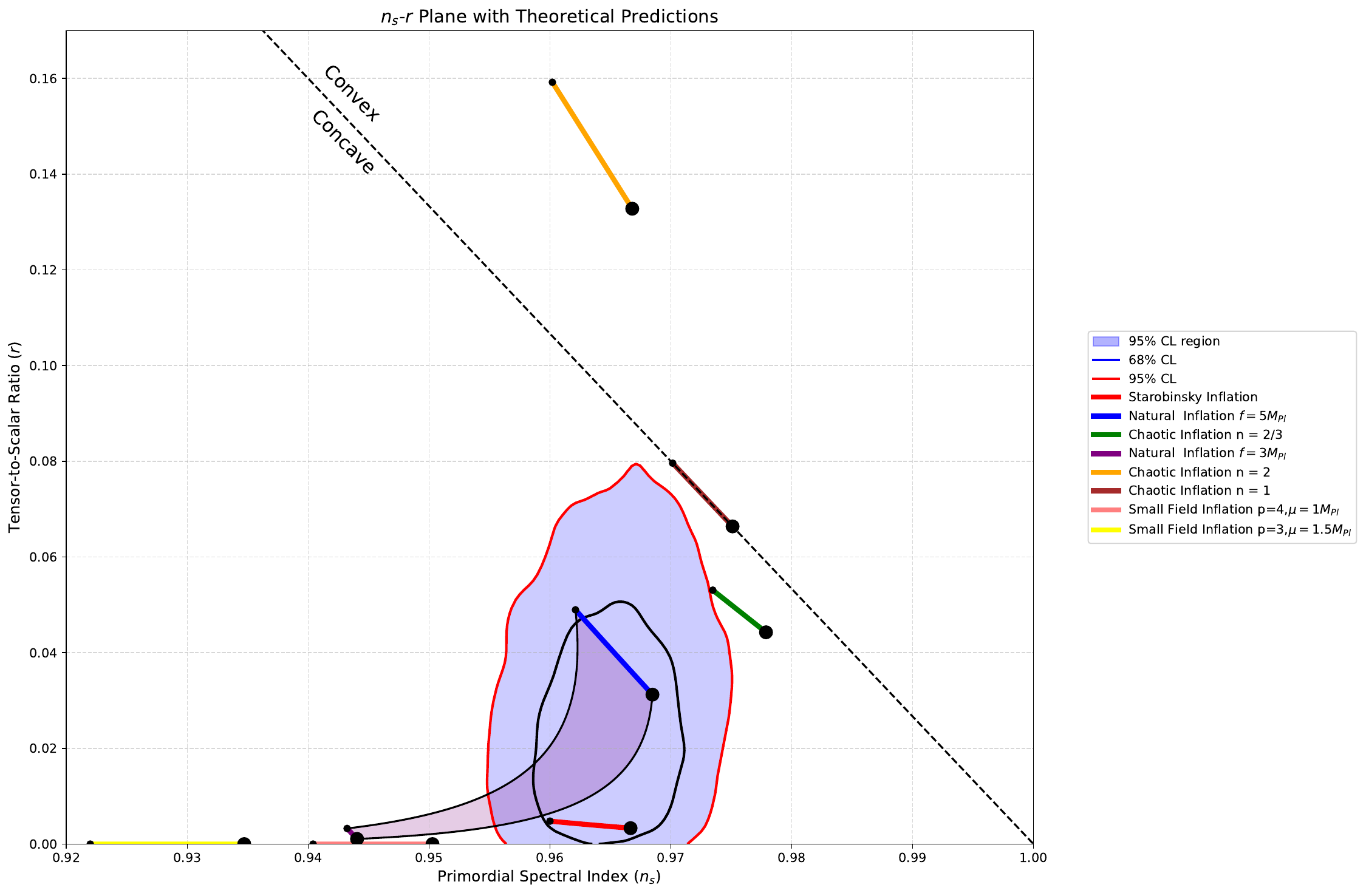 }  
    \caption{ The 68\% and 95\% confidence level regions for $n_s$ and $r$ of initial power spectrum, derived from Planck data in conjunction with BK15, compared to theoretical predictions from models above.}
\end{figure}

\section{Conclusion}

In summary, two essential parameters, $n_s$ and $r$ can be distinctly predicted in single-field inflation based on the principle of least action.  Potentials under various assumptions yield different results. Chaotic Inflation predicts a larger ratio than observations, while Small Field Inflation shows a small ratio accompanied by a deviation in the spectral index. Natural Inflation derives a suitable result when \(f \approx 5M_{\text{Pl}}\), but this introduces additional quantum gravitational effects. Starobinsky Inflation provides the best results among the four models.

Furthermore, the evolutions of the field also differ across the models. In Starobinsky Inflation and Chaotic Inflation, the magnitudes of the field decrease at an increasing rate, with the rate in Starobinsky Inflation increasing faster than that in Chaotic Inflation. In contrast, in the other two models, the magnitudes of the field increase throughout the entire inflationary stage.

\bibliographystyle{raa}
\bibliography{bibtex}
\end{document}